# Positive and negative design in stability and thermal adaptation of natural proteins

Igor N. Berezovsky, Konstantin B. Zeldovich, and Eugene I. Shakhnovich*

*Corresponding author

*Department of Chemistry and Chemical Biology, Harvard University, 12 Oxford St, Cambrigde, MA 02138*

## Abstract

The aim of this work is to elucidate how physical principles of protein design are reflected in natural sequences that evolved in response to the thermal conditions of the environment. Using an exactly solvable lattice model, we design sequences with selected thermal properties. Compositional analysis of designed model sequences and natural proteomes reveals a specific trend in  amino acid compositions in response to the requirement of stability at elevated environmental temperature - the increase of fractions of hydrophobic and charged amino acid residues at the expense of polar ones. We show that this "from both ends of hydrophobicity scale" trend is due to positive (to stabilize the native state) and negative (to destabilize misfolded states) components of protein design. Negative design strengthens specific repulsive nonnative interactions that appear in misfolded structures. A pressure to preserve specific repulsive interactions in non-native conformations may result in correlated mutations between amino acids which are far apart in the native state but may be in contact in misfolded conformations. Such correlated mutations are indeed found in TIM barrel and other proteins.

## Synopsis

What mechanisms does Nature use in her quest for thermophilic proteins? It is known that stability of a protein is determined, mainly, by energy gap – the difference in energy between native state and set of incorrectly folded (misfolded) conformations. Here we show that Nature makes thermophilic proteins by widening this gap *from both ends.* Energy of the native state of a protein is decreased by selecting strongly attractive amino acids at positions that are in contact in the native state (positive design). Simultaneously, energies of the misfolded conformations are increased by selection of strongly repulsive amino acids at positions which are distant in native structure; however these amino acids will interact repulsively in the misfolded conformations (negative design). These fundamental principles of protein design are manifested in "from both ends of hydrophobicity scale" trend observed in thermophilic adaptation, whereby proteomes of thermophilic proteins are enriched in extreme amino acids - hydrophobic and charged - at the expense of polar ones.  Hydrophobic amino acids contribute mostly to the positive design, while charged amino acids that repel each other in non-native conformations of proteins contribute to negative design. Our results provide guidance in rational design of proteins with selected thermal properties.

# Introduction

Despite recent advances in computational protein design [1], there is no complete understanding of basic principles that govern design and selection of naturally occurring proteins [2]. In particular the physical basis for the ability of proteins to achieve an adaptation to a wide variety of external conditions is still poorly understood. While several attempts to design proteins with a desired fold were successful [1,3], rational design of proteins with desired thermal properties is still an elusive goal. However, Nature apparently succeeds in doing so by ''designing'' proteins in hyperthermophiles that are stable and functional up to 110 C°. Thus in the absence of the complete solution of the protein design problem it is tempting to get clues from Nature as to how thermal properties of proteins can be modulated by proper sequence selection and which physical factors play a role in this process.

A clear manifestation of thermophilic adaptation can be found in a highly statistically significant variation of aminoacid compositions of proteomes between meso- and thermophilic organisms [4-7]. Recently, we showed that the total concentration of seven amino acids: I,V,Y,W,R,E,L is highly correlated with Optimal Growth Temperature (OGT) of an organism (R=0.93) [8]. The total concentration of IVYWREL combination of amino acids serves as a predictor of OGT with mean accuracy of 8.9 °C [8]. In this work we seek a fundamental theoretical explanation as to why Nature requires an elevated concentration of both hydrophobic and charged amino acids to design hyperthermostable proteins.

Our first goal here is to develop a minimalistic physical model of protein design that could help us to rationalize comparative proteomic analysis of thermo- and mesophiles. A crucial question is how to incorporate the environmental temperature in the model of protein design. Two factors may play a role. The first effect is due to fundamental statistical mechanics of proteins that posits that stable and foldable proteins should have an ''energy gap'' [9-12]. Specifically, the stability of the native state of a protein is determined by the Boltzmann factor $\exp(-\Delta E/k_B T)$, where $\Delta E$ is the energy gap between the native state and lowest energy completely misfolded structures [12-14] [15,16] Therefore in order to maintain their stability at elevated temperature the thermophilic proteins should have greater energy gap. In principle, the increase of energy gap can be achieved by lowering energy of the native state (positive design), raising energy of misfolds (negative design), or both. Another factor that may affect protein thermostability is a possible dependence of fundamental interactions (e.g. hydrophobic forces) on temperature. However, the temperature dependence of different types of interactions may be very complex, and it remains a subject of controversy as to how and to what extent it influences the stability of proteins [17-20]. Our approach to this complex issue is simple: consider first how far one can go based on purely statistical-mechanical analysis of protein thermostability without resorting to explanations based on temperature dependence of various interactions. Specifically, here we use the 27-mer cubic lattice model of proteins [21,22]. The model features 20 types of amino acids which interact when they are nearest neighbors on the lattice; interaction energy depends on types of amino acids involved. The potential is derived from known protein structures and is temperature-independent [23]. For this lattice model all compact conformations can be enumerated [21] and, therefore, exact statistical-mechanical analysis is possible. Previously, protein thermodynamics [9,24], folding [25,26], and evolution [14,27,28] were extensively studied by using this model. We simulate the process of thermal adaptation by the design of 27-mer sequences with selected (at a given environmental temperature $T_{env}$) thermal properties[15,16]. The algorithm of design (see Methods) makes simultaneous unrestricted search in both conformational, sequence, and amino acid composition spaces. In our analysis we will focus on the amino acid composition of designed sequences as a function of the environmental temperature and we will compare the model findings with amino acid trends in real proteomes. Our main result is that thermal adaptation utilizes both positive design and negative design and we show that by increasing the content of amino acids from both extremes of hydrophobicity scale thermostable proteins achieve exactly that goal – hydrophobic residues help with positive design while elevated concentration of charged residues helps to achieve stronger negative design. Further, we find an interesting and potentially important aspect of negative design: Similar to positive design that strengthens certain native interactions, negative design can make *specific nonnative interactions* strongly repulsive. This in turn may lead to emergence of correlated mutations between amino acids that are not in contact in native structure.

### Results

We design lattice model proteins with selected thermostability as a first step towards modeling thermal adaptation of organisms. There is a direct connection between OGT (environmental temperature, or $T_{env}$) of an organism and the melting temperature of its proteins [29,30]. We used the P-design procedure to create model 27-mer sequences that are stable at selected $T_{env}$ (see Methods). We designed sets of 5000 model proteins for each $T_{env}$ in the range $0.3<T_{env}<0.8$ in Miyazawa-Jernigan dimensionless units. The average melting temperature $<T_{melt}>$ of lattice proteins is strongly correlated with $T_{env}$ (see Figure S1 in Supplementary Information) suggesting that the P-design procedure does work. It provides model proteins with desired stability in response to the increase of environmental temperature. The dependence of $<T_{melt}>$ on $T_{env}$ is close to linear and qualitatively matches the empirical linear relationship, $T_{melt}=24.4+0.93\ T_{env}$, between the average living temperature of the organism and melting temperature of its proteins [29].

As expected, the amino acid composition of designed proteins does depend on $T_{env}$ for which they were designed. To quantify the differences between "low-temperature" and "high-temperature" amino acid compositions, we plotted temperature dependencies of the fractions of hydrophobic (LVWIFMPC), weak hydrophophobic and polar (AGNQSTHY), and charged (DEKR) amino acids for designed lattice proteins (Figure 1(a)) and natural (Figure 1(b)) proteomes. Figure 1(a) shows a significant increase of the amount of charged residues (red triangles) and slight increase of hydrophobic amino acids (green squares) at the expense of polar (black dots) ones. Remarkably the results shown in Figure1 suggest that increase of thermostability is accompanied by growth of amino acid content from both extremes of hydrophobicity scale, adding both charged and hydrophobic residues. This observation is further highlighted in Figure 2 that shows amino acid by amino acid, how compositions of model proteins with $T_{env}$ in designed model proteomes for all 20 amino acids, ranked by their hydrophobicity according to the Miyazawa-Jernigan set of interaction parameters (see

Methods and Figure S2ab in Supporting Information for more detailed explanation). Figure 2 clearly shows that addition of amino acids to thermophilic model proteomes occurs from the extremes of hydrophobicity scale while the middle is depressed. The content of charged (Asp, Glu, Lys, Arg; DEKR) and four of the hydrophobic (Ile, Leu, Phe, Cys; ILFC) residues is increased with temperature at the expense of other residues, mostly polar ones. This observation shows that combining amino acids with maximum variance in their hydrophobicity is crucial for creating hyperthermostable model proteins. We refer to this effect as to "from both ends of hydrophobicity scale" trend.

For comparison, we analyzed the variation of amino acid composition in fully sequenced bacterial proteomes (83 species in total, see complete list in Supporting Informaiton) of psycho-, meso, thermo, and hyperthermophilic prokaryotes (habitat temperatures from -10 to +110 ºC, see Table 1 in Supporting Information). Importantly, amino acid composition of 83 natural prokaryotic proteomes reveal similar trends, an increase of the contents of hydrophobic and charged residues and a decrease of the content of polar ones (Figure 1(b)). For a more direct comparison of the predictions of our model with the properties of natural proteomes, in Figure 3 we plotted the temperature derivative of the fraction of each of the amino acids in designed lattice proteins against the corresponding temperature derivative calculated over the 83 natural proteomes. The observed positive significant correlation (R=0.56, p=0.01) suggests that generic physical factors captured by this simple statistical-mechanical model played a major role in shaping the amino acid composition patterns across a wide range of habitat temperatures.

We hypothesize that the generic character of the "from both ends" trend that is universally observed in the model and in natural proteins is related to the positive and negative elements of design. In this case, one (hydrophobic) end of the scale is responsible for positive design while another (hydrophilic) end provides negative design. In order to test this hypothesis we first studied how the energy gap between the energy of the native state and that of misfolded conformations for the designed model proteins depends on $T_{env}$ (Figure 4). Positive design is the major contributor to the effect (the slope of the temperature-dependent energy decrease of the native state with growth of $T_{env}$ is -5.22; Figure 4, black line), while the increase of the average energy of decoys with $T_{env}$ (slope +1.64; Figure 4, orange line) is pronounced, but less significant. Nevertheless the results presented in Figure 4 provide clear evidence that negative design works, along with positive design, in the selection of thermostable model proteins.

The findings shown in Fig.4 demonstrate that indeed both positive and negative design act in enhancing thermostability of model proteins. However, the question remains as to how are positive and negative design related to ''from both end'' trend in amino acid compositions as shown in Figure 2? To address this question, we plot the number of contacts between amino acids whose content grows with $T_{env}$, according to Figure 2. Figure 5 shows how the average number of contacts (per structure) within both groups of aminoacids, FILC and DEKR, in native conformations and in misfolded decoys depends on $T_{env}$. Remarkably, we see that in decoy structures the growth of the number of contacts occurs only within the ''charged'' group DEKR, some of which according to the Miyazawa-Jernigan potential repel one another. On the other hand, the number of contacts in hydrophobic group in decoys does not change despite an overall increase of concentration of these amino acids in sequences designed at higher $T_{env}$. This result shows that while strongly mutually attractive hydrophobic groups provide lower energies of native states for hyperthermophilic model proteins, the growth in concentration of ''charged'' (DEKR) groups mainly contributes to negative design factor by raising average energy of misfolded conformations. Remarkably, the average number of contacts between hydrophobic groups (FILC) in misfolded conformations remains roughly the same in mesophilic and hyperthermophilic model proteins despite significant growth in overall concentration of these groups in hyperthermophiles. Therefore the data shown in Figure 5 indicates that ''from both ends'' trend in amino acid composition is directly related to positive and negative design in stabilization of hyperthemophilic model proteins.

The data presented so far provides insight into averaged (over many model proteins) contributions to the energies of native conformation and decoys. However a question arises whether negative design works by increasing ''average'' non-native interactions or by strengthening certain specific repulsive non-native interactions. Indeed, negative design may be based on introducing a few energetically disadvantageous nonnative contacts that are persistent in many decoy structures increasing their energy [2,31]. Therefore, nonnative contacts responsible for negative design may well be specific for each sequence, making this effect more detectable if individual proteins are considered.

The exact nature of the lattice model makes a detailed residue-by-residue analysis of the action of both positive and negative design possible. To this end it is instructive to identify interactions, native and nonnative contacts, between residues that play especially important role in stabilization of the native state and destabilization of decoys. The key idea here is that such important interactions should be conserved in all sequences that fold into a given structure. While identities of amino acids that form such a contact may vary from sequence to sequence, the strength (or energy) of key native or nonnative contacts will be preserved: it will be either strongly repulsive or strongly attractive for all sequences that fold into a given structure [32]. Therefore in order to identify such key contacts, distributions of energies of native and nonnative contacts in multiple sequences that fold into the same native structure should be considered. Such analysis can reveal not only conserved strong native contacts but also possible conserved strong repulsive nonnative contacts. To investigate such possibility, we designed 5000 lattice proteins that all fold into the same (randomly chosen) native structure. To achieve that we used the design algorithm similar to P-design (see Methods), but for a fixed native structure, and checked *a posteriori* that the target structure is indeed the native state for all 5000 sequences. We designed a set of 5000 ''mesophilic'' sequences at $T_{env}$=0.2 and 5000 "hyperthermophilic" sequences that fold to the same structure but are much more stable ($T_{env} = 0.8$).

The concept of native and nonnative contacts for our lattice model is illustrated in Figure 6a. It is a cartoon with a zoom-in into the contact matrix of the lattice structure used in simulations. The contact matrix of any compact lattice conformation contains all native (green, total 28 in any structure) and all possible nonnative (blue, total 128) contacts. All other contacts (red) are prohibited according to the properties of the cubic lattice. In order to identify important native and nonnative contacts whose energies are conserved we applied the following procedure. First, for each of the 5000 sequences that fold into selected structure we calculated energies of 28 native and 128 possible nonnative contacts in this structure (using the identities of residues and Myazawa-Jernigan potentials that were employed to design sequences). Next, for each contact we calculated average energy and its standard deviation over all 5000 designed sequences (see Figure 6b for illustration of this calculation). Contacts whose energy shows a very low standard deviation over all designed sequences are apparently the ones that are most important for stability. This procedure was carried out both for mesophilic sequences ($T_{env}$ =0.2) and for thermophilic sequences ($T_{env}$ =0.8) and the results are shown in Figure 7, which presents standard deviation of interaction energies of each native and nonnative contact over all 5000 designed mesophilic sequences (a) and hyperthermophilic sequences (b), plotted against the average (over 5000 designed sequences) energy of that contact. The plot consists of 28+128=156 points, covering all native and all possible nonnative interactions. The native state clearly defines conserved low- and high-energy native contacts (shown in black) in most of the sequences, as the standard deviation is the lowest at the extreme values of the energy. Conserved attractive interactions are in the protein interior, corresponding to the lattice analog of the hydrophobic core; apparently, they emerge due to the action of positive design. The nonnative contacts (red dots) follow a different pattern, with only a few conserved attractive interactions, suggesting the diversity of decoy structures. What is surprising to see, however, is that energies of certain *most repulsive* (high-energy) nonnative contacts show a very low standard deviation - indicating that such contacts may be as important for protein stability as conserved native ones Comparison of mesophilic and hyperthermophilic sequences shows clearly that emergence of strong and conserved   attractive and repulsive interactions in key native and nonnative contacts is directly related to sequence design that generates stable sequences: design of hyperthermostable sequences (Figure.7b) results in stronger and more conserved (lower dispersion of energy) attractive and repulsive specific native and nonnative interactions. The only reason that repulsive energies of nonnative contacts are conserved is that such contacts persist in certain frequent decoy structures and contribute to the widening of the gap between the native state and decoys. Such repulsive contacts are indirectly (via the sequence) related to a particular native state, are not numerous, and their role may be completely obscured in a "high-throughput" analysis where sequences with different native states are considered together, as in Figure 4. Therefore, we conclude that negative design involves a very specific strategic placement of repulsive contacts in certain decoy structures.

The results and analysis presented in Figure 7 have very important implication for real proteins. The requirement to conserve energy of key contacts in multiple sequences that fold into the same structure implies that amino acids forming such contacts can mutate in correlated way, for example by swaps. The observation that mutations may often occur as swaps to preserve specific attractive native and specific repulsive nonnative interactions leads to a prediction of a peculiar dependence between frequency of amino acid substitutions (as in e.g. in BLOSUM matrices [33]) and interaction energy between amino acids. Indeed, as illustrated in Figure 8, a correlated mutation in the form of a swap can manifest itself *in sequence alignment* as a substitution between amino acids that are making the swap. The implication is that frequent substitutions will be observed between amino acids that strongly attract each other (to preserve specific stabilizing native contacts). More interestingly and perhaps more surprisingly, frequent substitution are also predicted between amino acids that strongly repel each other (to preserve specific nonnative repulsive contacts). In other words, we predict that the scatter plot between elements of amino acid interaction energy matrix and substitution matrix will be non-monotonic with maxima at both extremes.

We tested this prediction by plotting the dependence of elements of substitution matrix BLOSUM62 [33] for 190 pairs of amino acids (synonymous substitutions are excluded) versus their interaction energy as approximated by the knowledge-based Miyazawa-Jernigan potential [23] (Figure 9). This analysis indeed reveals a non-monotonic shape: The parabolic fit in Figure 9a highlights highly significant non-monotonic nature of the dependence. The striking feature of this dependence is that most frequent substitutions are observed not only between most attractive amino acids but also between most repulsive ones. One could argue however, that high frequency of substitutions between amino acids that repel each other may be a trivial consequence of conserved substitutions that preserve the charge (R to K and E to D). However a detailed inspection of the upper right part of the plot in Figure 9a shows that this is not the case (Figure 9b). Indeed, frequent substitutions are observed between mutually repulsive amino acids with vastly different physical-chemical properties and encoded by very dissimilar codons such as, e.g. Serine to Asparagine, Glutamic Acid to Arginine etc. Several highly non-conservative substitutions show about ''random'' frequency (element of BLOSUM matrix close to zero, e.g. for Asn to Lys) but this may be due to compensation of two opposite effects: suppression of highly non-conservative substitutions (e.g. that change charge) and facilitation of correlated substitutions like the ones in the form of swaps as illustrated here.

Use of correlated mutations as predictors of spatial proximity of amino acids in the native structure has been proposed by many authors [34-37]. Indeed, statistical analysis shows that overall correlation between distance between amino acids and degree of correlation in multiple sequence alignments does exist [38]. However it has been noted that sometimes correlated mutations are observed between amino acids that are distant in native structure [34,39]. While sometimes such observations are discarded as false positives in the prediction algorithm [34], our analysis predicts that indeed residues that are distant in structure but may form important repulsive contacts in misfolded conformations may exhibit correlated mutations as illustrated in Figures 8 and 9.

As an illustration of the significance of correlated mutations between amino acids that are far apart in structure we consider a Tim-barrel fold protein triosephosphate isomerase (PDB id 7tim). Guided by the results of statistical analysis shown in Figure 9, we looked for pairs of residues with strong repulsion according to the Miyazawa-Jernigan potential, random or higher substitution rate between these residues according to the BLOSUM matrix, and highly correlated substitutions of these residues in two positions of the protein sequence in multiple sequence alignment for 7tim (see Methods). These residues should be not in contact in the native state and should not be involved in the functional site or into the protein-protein interactions in order to distinguish the effect that we seek from functional conservation.

We found correlated substitutions in the sequence of TIM barrel fold (7tim, chain a), according to the physicochemical characteristic, hydropathy [40], by using CRASP program which "estimates the contribution of the coordinated substitutions to invariance or variability of integral protein physicochemical characteristics" [41]. Four pairs of residues (Table 1) which have highly coordinated substitutions and repel each other according to the Miyazawa-Jernigan energy matrix were selected None of those residues belong to the functional site of triosephosphate isomerase, and the protein itself is a single-domain protein not involved into protein-protein interactions [42]. Four pairs of polar and charged residues and pair of charged residues were identified (see Figure 10a and Table 1). The shortest contact distance (C$\alpha$-C$\alpha$, 8.7 Å) is between charged Lys(84) and polar Gln(119), which excludes stabilizing interaction between them in the native structure. We found that correlated mutations between some of these amino acids occur as swaps in TIM barrel fold, possibly accompanied by conservative mutation. *E.g.* surface Lys 120 and Gln 85 in 7tim swap to Gln 120 Lys 85 in *T. maritima* thermophilc ortholog of triosephosphate isomerase (pdb id 1b9b). Even more strikingly Gln85, Lys 213 pair in 7tim (distance in native structure 30 Å) is replaced by Lys 85, Asn 213 in 1b9b (Figure 10b). This pair of residues shows highly correlated substitution pattern in TIM-barrel multiple sequence alignment despite the fact that these are very distinct amino acids.

## Discussion

Stabilization of thermophilic proteins is achieved by negative and positive design working together, i.e. the gap ''opens'' from both sides: decreasing energy of the native state and at the same time increasing the energy of misfolded conformations. This factor is responsible for the ''from both ends of hydrophobicity scale'' trend observed in model and real [8] thermophilic proteomes. In particular, our recent analysis of complete bacterial proteomes [8] revealed that proteomes of thermophilic bacteria are enriched in both hydrophobic residues (IVYLW) and charged ones (ER) while all polar residues are suppressed. Discrepancies between different hydrophobicity scales [43], the statistical nature of knowledge-based Miyazawa-Jernigan potential [23], and limitations of the lattice model make it impossible to quantitatively compare the content of individual amino acids in lattice and natural proteomes or exactly predict the amino acid composition of thermophilic proteomes with very high accuracy from lattice model calculations. Nevertheless our lattice calculations are in semi-quantitative agreement with data on natural proteomes, (see. Figures 1 and 3) and exhibit the same ''from both ends of hydrophobicity scale'' trend in amino acid composition adaptation in response to elevated habitat temperature.

The knowledge-based Miyazawa-Jernigan potentials, derived from native structures of proteins are certainly a crude approximation to real protein energetics [44]. A question arises as to whether our observations are generic or are they due to the specific potential used to design model proteins? A detailed comparison of several potentials – all atom and group-based derived by different methods was carried out recently in our lab [45] . Remarkably we found that despite differences in detail all these potentials reflect the same dominant contributions to protein stabilization. It appears that dominant contributions to energy gaps in proteins come principally from two types of interactions – hydrophobic interactions and electrostatics [45]. Further it was found that knowledge-based potentials derived using structures of mesophilic and thermophilic proteins are virtually indistinguishable (KZ and ES, unpublished results). These observations suggest that "from both ends of hydrophobicity scale" trend. observed in model calculations and in real proteome is a robust phenomenon, reflecting basic physical principles of protein design, rather than a consequence of a specific potential set used in calculations.

While positive design [46] is universally used in experiment, the role and omnipresence of negative design are still under discussion [47]. The main challenge in the study of negative design stems from the difficulties in the modeling of relevant misfolded conformations and energetic effects of mutations that destabilize them [47]. It was shown that charged residues can be effectively used in negative design [31]. Another indirect evidence of the contribution of charged residues to negative design emerges from site-directed mutagenesis, where mutations of polar groups to charged ones on the surface of a protein leads to protein stabilization even in the absence of salt-bridge partners of the mutated group [48-50]. It was shown in a series of experiments [48,49,51], that surface electrostatic interactions provide a marginal contribution to stability of the native structure, hence their possible importance for making unfavorable high-energy contact in decoys. An alternative view, proposed recently by Makhatadze and coauthors, suggests that long-range electrostatic interactions may contribute to stability of the native state [52]. However at normal physiological conditions the range of electrostatic interactions is limited due to Debye screening and hardly exceeds 8A. Our simulations and proteomic analysis point out to a possible role of some surface charged residues as contributing to destabilization of misfolded structures through negative design mechanism.

Positive and negative elements of design affect the evolution of protein sequences. The dependence of substitution rates in sequences of natural proteins (BLOSUM62 substitution matrix) on interaction energies according to knowledge-based

Miyazawa-Jernigan potential has a peculiar non-monotonic shape showing elevated substitution rates between residues that attract each other as well as between residues that repel each other. The physical reason for this phenomenon is the same as for ''both ends of hydrophobicity scale'' trend: simultaneous action of positive and negative design. Upon substitutions, energy of attractive contacts in native states should be preserved as well as energies of specific repulsive contacts in misfolded conformations. Apparently both these factors act in concert to preserve energy gap in proteins.

Our study deepens an understanding of correlated mutations in proteins. With regards to native contacts, the fact that amino acids making strongly attractive native interactions should exhibit correlated mutations had been realized long ago. Several authors proposed to use correlated mutations as a tool to determine possible native contacts from multiple sequence alignment [34-37]. However this suggestion is complicated by the observation that correlated mutations are often found between residues that have no obvious functional role and are distant in structure: [34] [39,53,54]. Using double mutant technique, Horovitz and coauthors [35] suggested a relation between correlated mutations and energetic connectivity (i.e. nonadditivity of stability effects in double mutation cycles) between corresponding amino acids. Green and Shortle [55] showed that amino acids that are distant in structure may indeed be ''energetically coupled'', attributing this effect to influence of mutations on unfolded state of proteins, consistent with our findings. Lockless and Ranganathan [39] suggested that a ''pathway of energetic connectivity'' exists between distant residues that exhibit correlated mutations. Fodor and Aldrich [38], however, examined several other proteins and argued against the "general principle of isolated pathways of evolutionary conserved energetic connectivity in proteins''. Here we show that negative design that destabilizes misfolded conformations of proteins may be responsible for correlated mutations between residues that are far apart in native structures

In this work we developed a simple exact model of thermophilic adaptation and discovered fundamental statistical-mechanical rules that Nature uses in her quest to enhance protein stability. While many other factors, including dependence of hydrophobic and other interactions on temperature, certainly play a role in protein stabilization, the action of positive and negative design found and described here in a minimalistic model appears to be a basic universal principle determining evolution of sequences of thermostable proteins. A better understanding of fundamental principles of protein design and stability makes it possible to decipher peculiar signals that emerge in the analysis of mesophilic and thermophilic genomes and proteomes [8] and in many studies of correlated mutations in proteins [34,36,54].

## Methods

We use the standard lattice model of proteins as compact 27-unit polymers on a 3x3x3 lattice [21].The residues interact with each other via the Miyazawa-Jernigan pairwise contact potential [23]. It is possible to calculate the energy of a sequence in each of the 103346 compact conformations allowed by the 3x3x3 lattice, and the Boltzmann probability of being in the lowest energy (native) conformation,

$$P_{nat}(T_{env}) = \frac{e^{-E_0/T_{env}}}{\sum_{i=0}^{103345} e^{-E_i/T_{env}}},$$

where $E_0$ is the lowest energy among the 103346 conformations, and $T_{env}$ is the environmental temperature. The melting temperature $T_{melt}$ is found numerically from the condition $P_{nat}(T_{melt})$=0.5. Note that if the energy spectrum $E_i$ is sparse enough at low energies, the value of $P_{nat}$ is determined chiefly by the energy gap $E_1$-$E_0$ between the native state and the closest decoy structure that has no structural relation to the native state.

To design lattice proteins, we use here a Monte-Carlo procedure (P-design, [15,16]) that maximizes the Boltzmann probability $P_{nat}$ of the native state by introducing mutations in the amino acid sequence and accepting or rejecting them according to the Metropolis criterion. As this procedure takes the environmental temperature $T_{env}$ as an input physical parameter, and generates amino acid sequences designed to be stable at $T_{env}$, it is an obvious choice for modeling the thermophilic adaptation.

Initially, the sequence is chosen at random; the frequencies of all amino acid residues in the initial sequences are equal to 5 percent. At each Monte-Carlo step, a random mutation of one amino acid in a sequence is attempted, and $P_{nat}$ of the mutated protein is determined. The native structure is determined at every step of the simulation; generally the native state changes upon mutation of the sequence. If the value of $P_{nat}$ increased, the mutation is always accepted; if $P_{nat}$ decreased, the mutation is accepted with the probability exp[-($P_{nat}$(old)-$P_{nat}$(new))/$p$], with $p$=0.05 (a Metropolis-like criterion). We chose p=0.05 so that the average melting temperature of designed proteins is higher than the environmental temperature (see Figure S2 in Supporting Information), in agreement with experimental observations [30,56]. The design procedure is stopped after 2000 Monte-Carlo iterations. Such length of design runs is sufficient to overcome any possible effects of the initial composition of the sequences, so the amino acid composition of the designed sequences depends only on the environmental temperature $T_{env}$.

To relate the trends in amino acid composition with the physical properties and interaction energies of individual amino acids, we use hydrophobicity as a generic parameter characterizing an amino acid [43]. To characterize the hydrophobicity of amino acids in the simulations, we make use of the fact that the Miyazawa-Jernigan interaction energy matrix is very well approximated by its spectral decomposition [44]. Interestingly, it is sufficient to use only one eigenvector **q**, corresponding to the largest eigenvalue, so the interaction (contact) energy $E_{ij}$ between amino acids of types $i$ and $j$ reads $E_{ij} \approx E_0 + \lambda q_i q_j$ [44]. In this representation, hydrophobic residues have the largest values of $q$, while hydrophilic (charged) residues correspond to small $q$.

All sequences of TIM barrel folds with length less than 300 amino acid residues were extracted according to the SCOP database description [57]. Identical sequences were excluded from further consideration. Remaining sequences (total 39) were aligned against the sequence of the triosephosphate isomerase (7tim.pdb, chain a) by using Kalign web-server for multiple alignment of protein sequences (http://msa.cgb.ki.se/cgi-bin/msa.cgi, [58]).

Correlated substitutions in the multiple alignments were determined by using CRASP program (http://wwwmgs.bionet.nsc.ru/mgs/programs/crasp, [41]). The CRASP program gives the correlation coefficient between the values of physicochemical parameters at a pair of positions of sequence alignment. We chose hydropathy [40], as a physicochemical characteristic appropriate for establishing correlated mutations of interest. Only significant correlations, with correlation coefficient higher than critical threshold (0.311) were considered.

The complete genomes were downloaded from the NCBI Genome database at http://www.ncbi.nih.gov/entrez/query.fcgi?db=Genome (see Table S1 in Supporting Information).

## Acknowledgments

This work is supported by NIH. INB was supported by the Merck Fellowship for Genome-related research. We thank George Makhatadze for useful correspondence.

## Supplementary Figure Captions

**Figure S1.** Average melting temperature of designed lattice proteomes (5000 sequences each) depending on the environmental temperature $T_{env}$ entering the P-design procedure (2000 Monte Carlo mutation steps to generate each sequence).

**Figure S2.** Temperature dependences of amino acid fraction for Val (**a**) and Glu (**b**) in 204 natural psycho-, meso, thermo, and hyperthermophilic proteomes (habitat temperatures from -10 to +110 ºC, see Table S1 for optimal growth temperatures and references).

## Supplementary Table Legend

Prokaryotes with completely sequenced genomes and their optimal growth temperatures. The columns are: **NN**, number; **Organism**, name of the organism; **OGT**, optimal growth temperature, °C; **Source**, source of the optimal growth temperature.

**ATCC**: American Type Culture Collection, accessed at http://www.atcc.org

**DSMZ**: German Collection of Microorganisms and Cell Cultures,

accessed at http://www.dsmz.de

 **PGTdb**: The Prokaryotic Growth Temperature Database, accessed at

http://pgtdb.csie.ncu.edu.tw

Reference: S. L. Huang, L. C. Wu, H. K. Laing, K.T. Pan, and J. T. Horng,

*Bioinformatics*, Vol. 20, pp. 276-278, 2004.

The complete genomes were downloaded from the NCBI Genome database at

http://www.ncbi.nih.gov/entrez/query.fcgi?db=Genome

**Table 1. Most significantly correlated substitutions in triosephosphate isomerase (7tim, chain a).**

| Pair of amino acid residues | Correlation coefficient, R | Distance in the structure, Å |
|---|---|---|
| Lys(84) – Gln(119) | 0.33 | 8.7 |
| Lys(84) – Asn(213) | 0.51 | 29.3 |
| Gln(119) – Glu(133) | 0.42 | 33.9 |
| Lys(134) – Ser(202) | 0.32 | 26.1 |

**Figure legends**

**Figure 1** Temperature dependences of the fractions of hydrophobic (LVWIFMPC), weak hydrophophobic and polar (AGNQSTHY), and charged (DEKR) amino acids plotted against temperature $T_{env}$ in the design experiment **(A)** and for real proteomes **(B)**. The data received in P-design procedure applied to sets of 5000 27-mer sequences with random amino acid composition, using a different value of $T_{env}$ for each set, $0.3 < T_{env} < 0.8$ ($T_{env}$ is measured Miyazawa-Jernigan dimensionless units). Temperature dependences of the fractions of amino acids in natural prokaryotic proteomes is plotted in **(B)** against optimal growth temperature (OGT) of the organism. There are total 83 natural proteomes with optimal growth temperatures spanning interval from -10 to +110 ℃.

**Figure 2.** Temperature derivatives of the fractions of amino acids plotted against their hydrophobicity $q$ (see Methods for definition of hydrophobicity parameter q in Miyazawa-Jernigan parameter set). The temperature derivatives were obtained as slopes from linear regression between the frequency of every of the 20 amino acids in the designed proteome and $T_{env}$ for which these sequences were designed The parabolic fit (in red) is to guide the eye.

**Figure 3.** The scatter plot between the temperature derivatives of the fraction of each of 20 amino acids in designed lattice proteins (y-axis) against the corresponding temperature derivative calculated over the 83 natural proteomes (x-axis). The correlation coefficient is R=0.56.

**Figure 4.** Temperature dependence of the contributions of positive and negative design to the gap in simulations of lattice model sequences designed to be stable at various temperatures. $T_{env}$. 5000 sequences were designed at each $T_{env}$ Positive design results in significant lowering of the native state energy (energy decrease in the interval of temperatures $T_{env}$=0.2÷0.8 has a slope equal to -5.22, black line). Temperature-dependent increase of the decoys' energy is less pronounced, pointing to the specific nature of the negative design. Slopes of the temperature dependences are: -3.57 (blue line) for the first decoy; 1.64 (orange line ) for the interaction energies averaged over all decoy structures, and 3.30 (red line) for the maximal energy structure.

**Figure 5.** Average number of contacts between amino acids whose fraction increases in thermophilic model sequences (see Fig.1) **(A).** Contacts in native structures. Contacts between ''charged'' (DEKR) residues are shown in red and contacts between ''hydrophobic'' (CFIL) residues are shown in black. **(B).** Misfolded structures. Color coding is the same as in (a).

**Figure 6**. Illustration of the calculation of energy dispersion of native and non-native contacts for the lattice model. **(A)**. Zoom-in into contact matrix of one of the 103346 compact lattice conformations, the cartoon. Even/even, odd/odd, diagonal, (i,i+1), (i,i+2) contacts do not exist in a 3x3x3 lattice (red). Every compact conformation has 28 contacts considered native for this conformation (green). There are also total 128 contacts that may appear inj alternative conformations, they serve as nonnative contacts for this conformation (shown in blue). **(B)** Calculation of average energies and dispersion of all native and nonnative contacts – illustrative example. 10 aligned sequences all folding into the same shown structure are presented. An example of residues making native (cyan)

and nonnative (red) contacts is shown. As an illustration energy of the selected native and non-native contacts are shown for each sequence (this energy is calculated according to the identities of amino acids forming a contact using Miyazawa-Jernigan knowledge-based potential) Average energy over all 10 aligned sequences of shown native and non-native contacts and its standard deviation are calculated for illustration here.

**Figure 7.** Average interaction energies and their standard deviations for all native (black dots) and nonnative (red triangles) contacts calculated over 5000 sequences having the same native state. **(A)**. design of "mesophilic" lattice proteins, $T_{env}$=0.2; **(B).** design of "hyperthermophilic" lattice proteins, $T_{env}$=0.8.

**Figure 8.** Schematic illustration of the concept of mutations by swaps**. (A)** A cartoon schematically shows how mutations by swaps preserve the contact energy for some important native (blue) and nonnative (red) contacts in Sequence 2 which folds to the same native state as Sequence 1. Higher energy misfolded structures are also shown schematically for both sequences. Swap of ILE and VAL in Sequence 2 does not change energy of the native contact between these amino acids in native structure.. Repulsion between ARG and LYS residues is also preserved in decoy structure in Sequence 2 if ARG and LYS swap their positions in this sequence as compared to Sequence 1. **(B)** Implication of swaps for multiple sequence alignments. Residues that swap in structure appear as substitutions in a multiple sequence alignment.

**Figure 9.** Scatter plots showing the dependence of the elements of the BLOSUM62 substitution matrix on interaction energies between amino acid residues (approximated by Miyazawa-Jernigan parameters [23]). Only non-synonymous substitutions are presented. **(A)** A complete plot showing all 190 possible non-synonymous pairs. Red lines represent

parabolic fit to highlight non-monotonic nature of the plot. $R^2=0.36$, $p<10^{-4}$ for the fit and the coefficient at $X^2$ is $0.18\pm0.02$. An alternative linear fit (not shown) is highly insignificant: $R^2=0.01$, $p=0.89$ **(B)** Blow-up of the right upper corner of **(a)** with amino acid pairs labeled.

**Figure 10.** Correlated mutations and swaps in representatives of the TIM barrel fold. **A.** Four pairs of surface residues exhibiting correlated mutations in triosephosphate isomerases from *S. cerevisiae* (7tim, chain a): Lys84(red) − Gln119(red), Lys84(red) − Asn213(cyan), Gln119(red) − Glu133(orange), and Lys134(green) − Ser202(green). **B.** Comparison of *S. cerevisiae* triosephosphate isomerase with triosephosphate isomerase from *T. maritima* (shown here: 1b9b, chain a) reveals two swaps of surface amino acid involved into correlated mutations in triosephosphate isomerase: Gln85(red) − Lys120(red) and Gln85(red) − Lys(213). These amino acids do not interact in native structure of either molecule. Positions are numbered according to the 1b9b sequence aligned with 7tim.

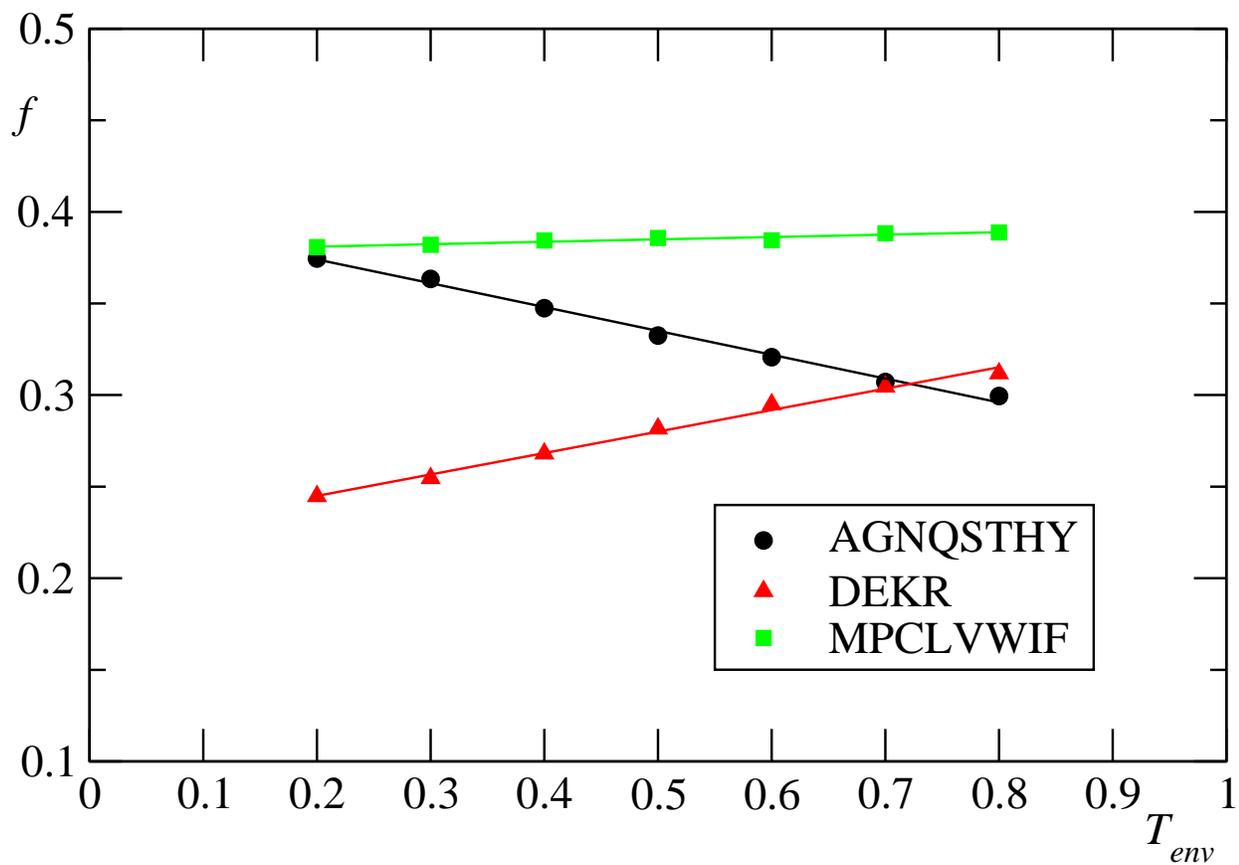

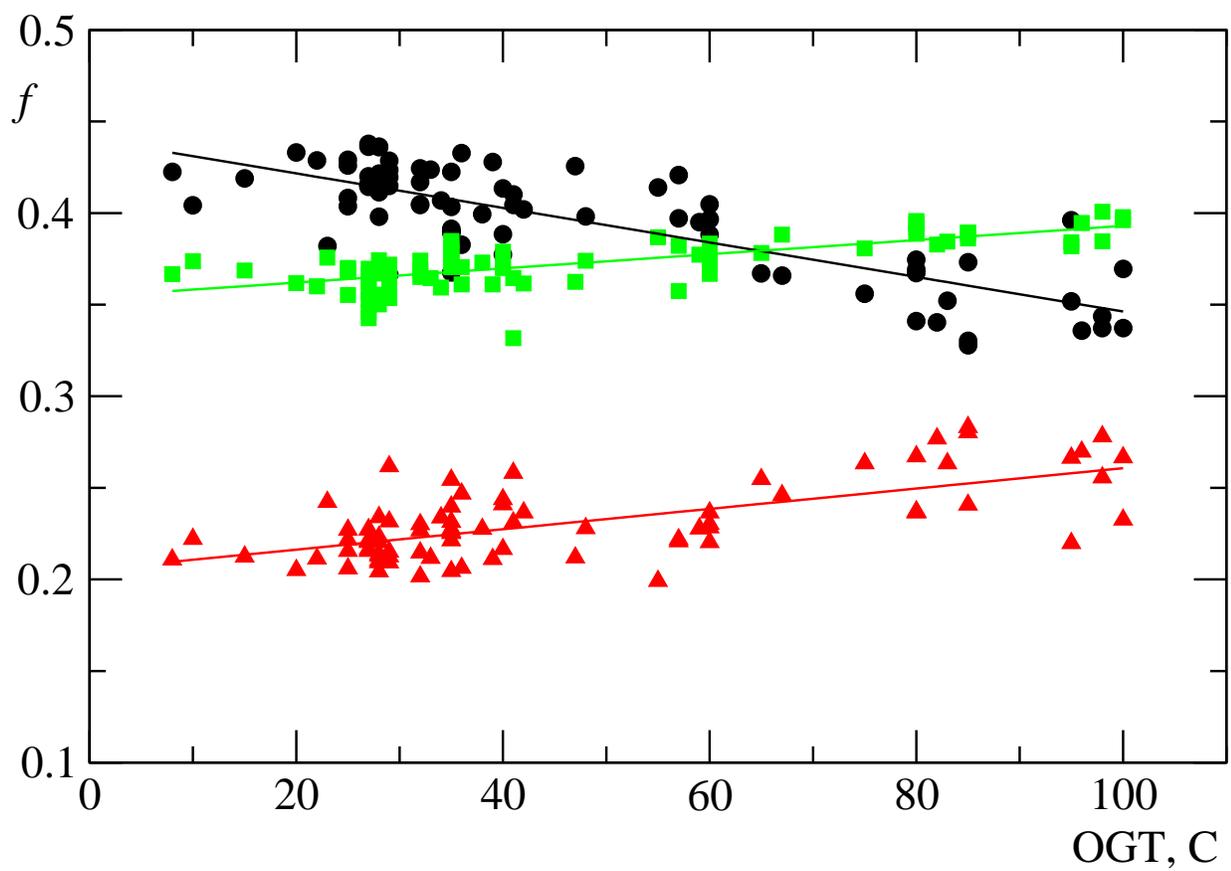

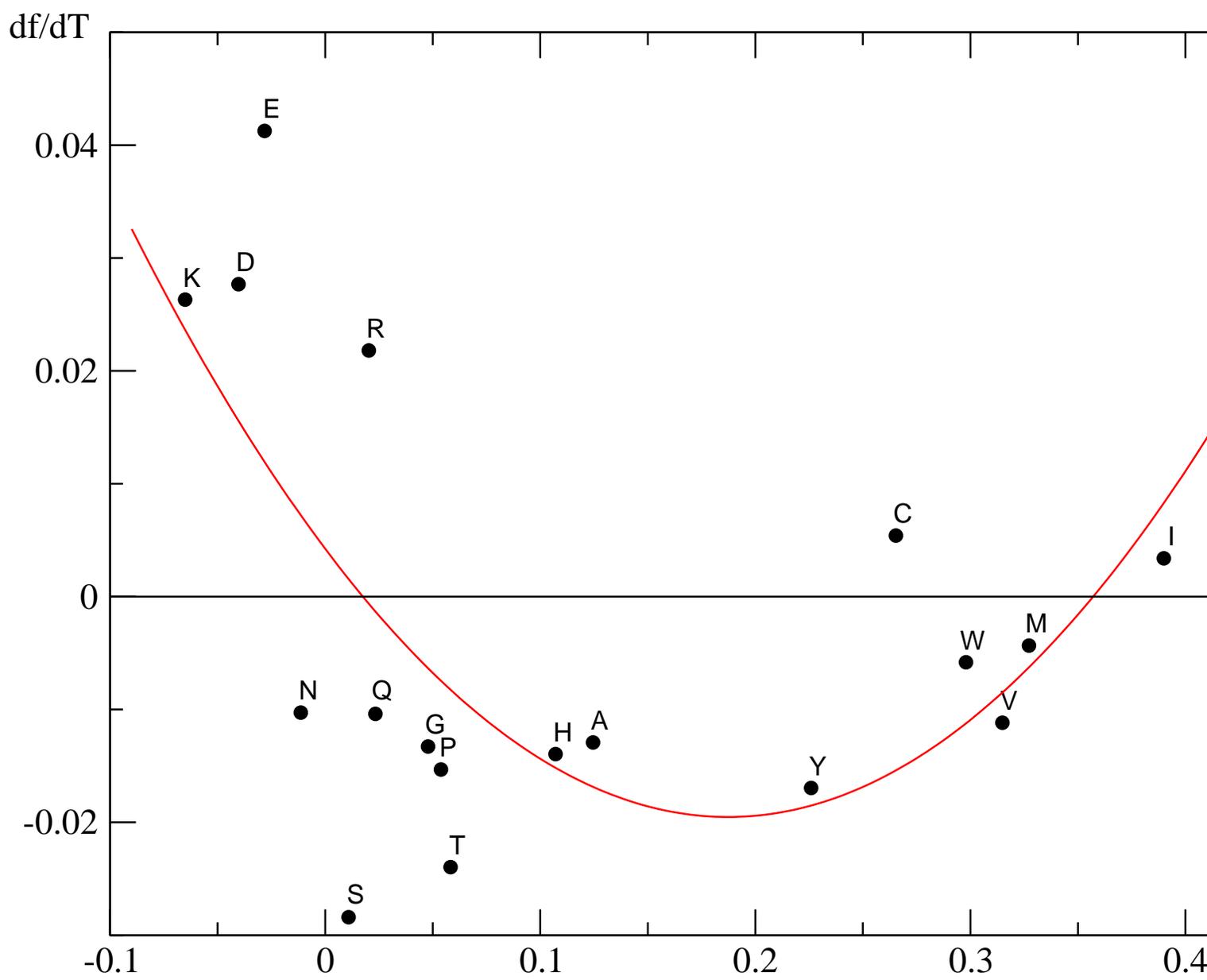

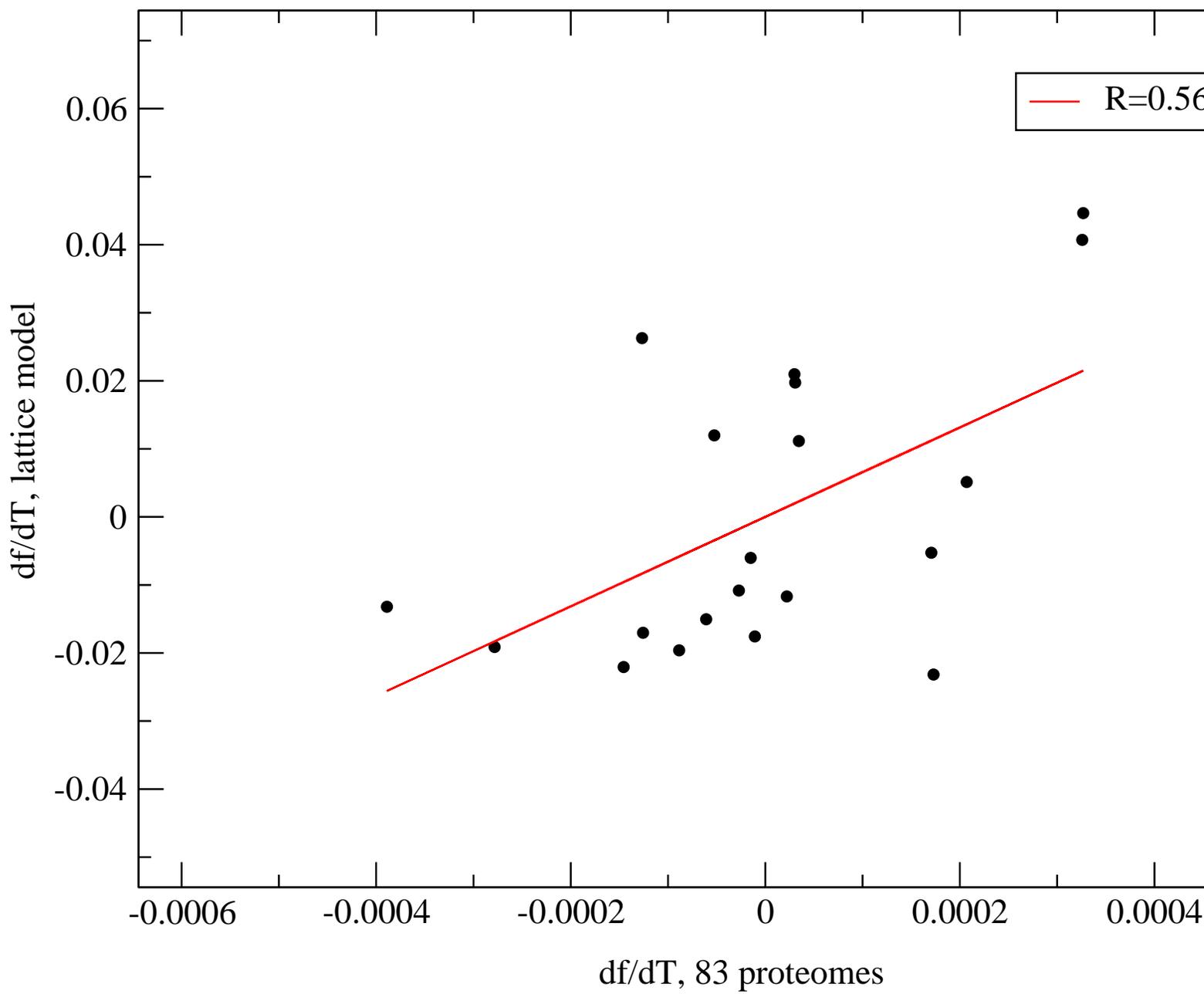

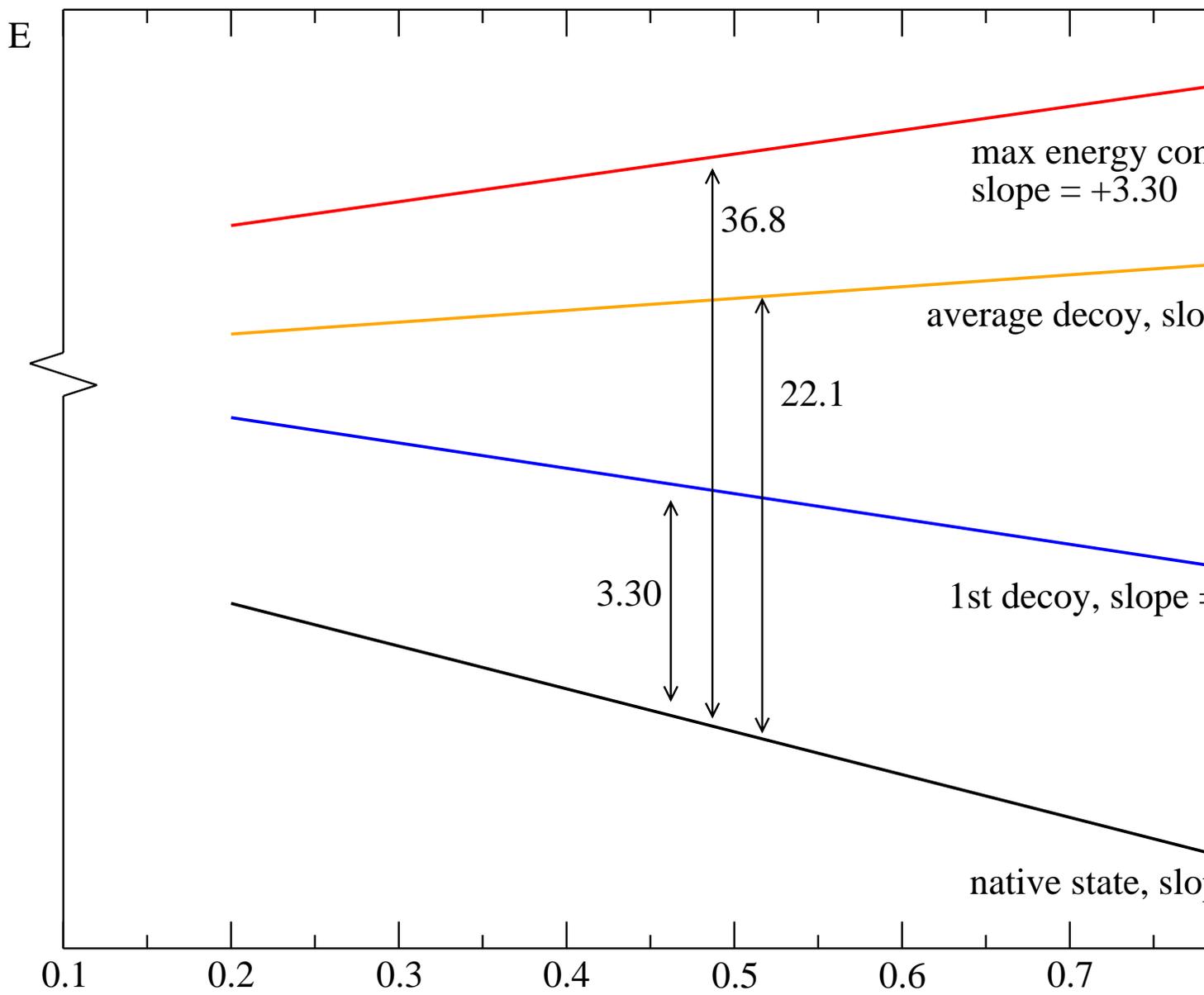

native contacts per chain

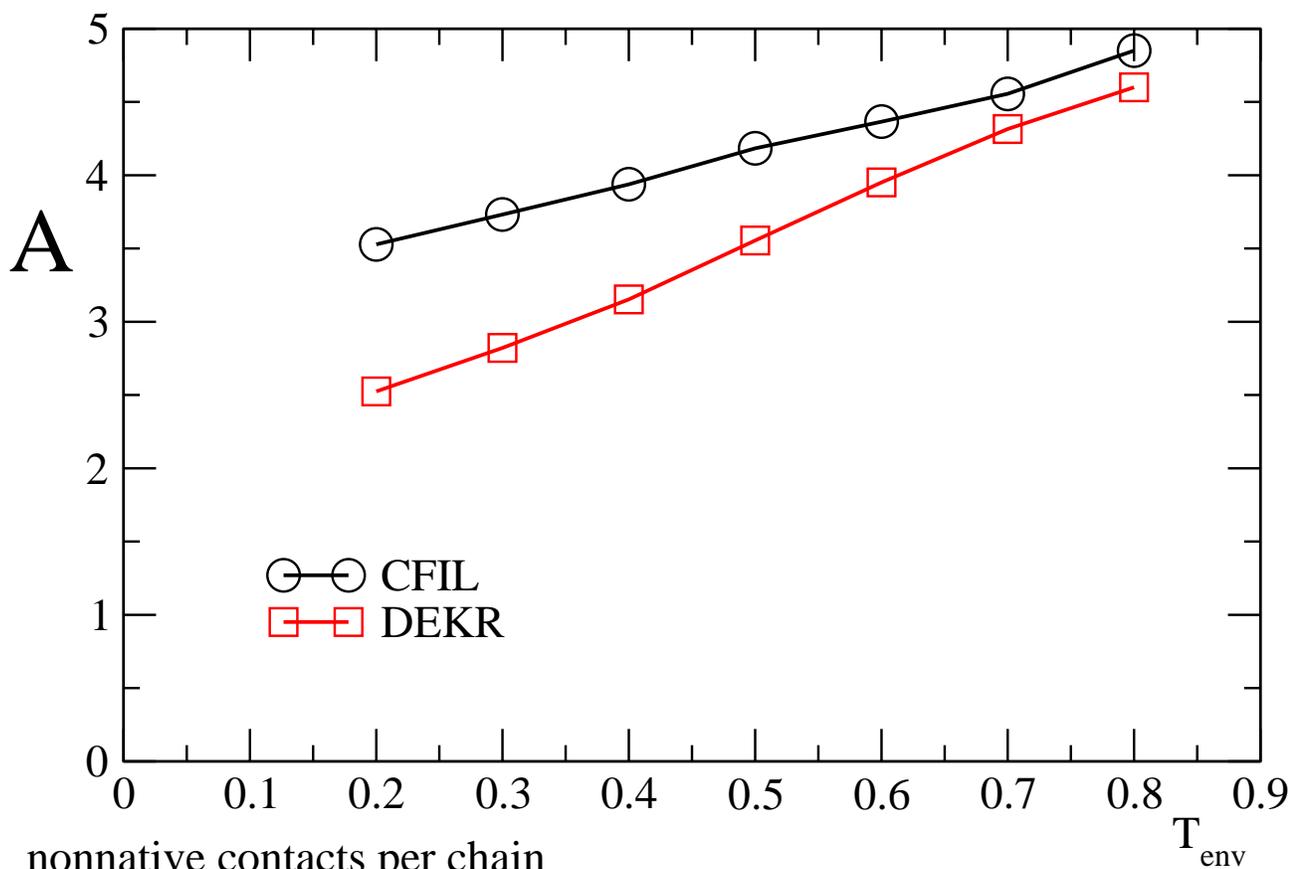

nonnative contacts per chain

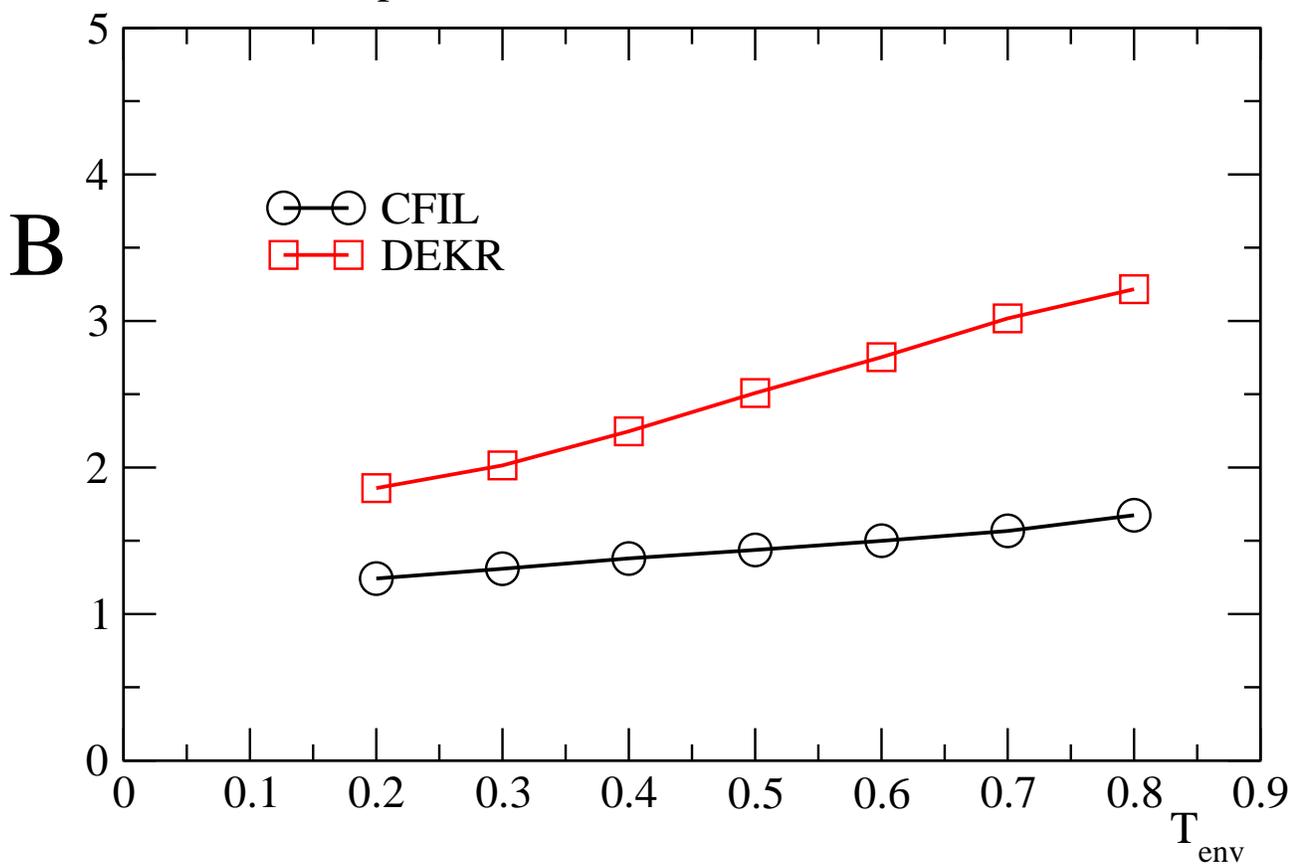

**A**

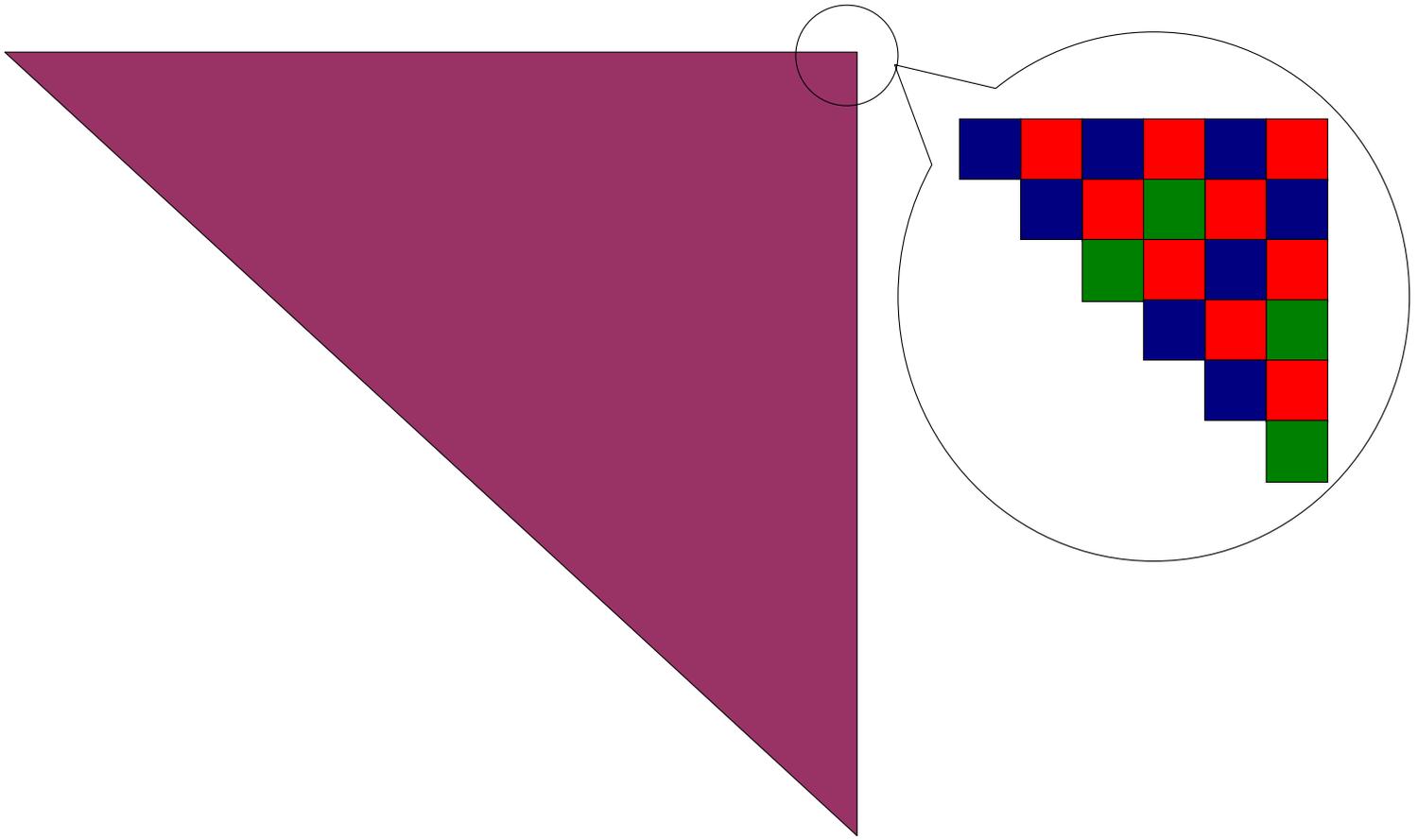

**B**

| Sequence | Native contact energy | Nonnative contact energy |
|---|---|---|
| PCRLRRYRPPMVEGNDYNACRQPYCVM | -4.62 | -1.64 |
| SCKPHEYGSRFNSEKNWELCQIECCWW | -5.06 | -1.80 |
| DDEWPKCEQWTSKCGNITLKHTPKDIM | -6.54 | -1.74 |
| RLHIASMQRKWRRMRLIEMFQYPALLY | -7.04 | -2.15 |
| KLRPTPCHHRFGPCECMPCAELEAIHY | -3.98 | -1.70 |
| DRGQTCYYGVCVSKDRMHLSTRACHFC | -6.56 | -2.15 |
| DQCCNSYDSWFVNIQEMTIRAYEVYMC | -5.46 | -3.11 |
| KMPSRGRKPGTVEVREFDVYKYSTLIG | -6.84 | -1.33 |
| DENVECYDPVLMKQNHWGFHGWVMHFV | -6.16 | 1.74 |
| | | |
| **Average contact energy, <*E*>** | -5.81 | -1.92 |
| **Standard deviation, σ_E** | 1.07 | 0.51 |

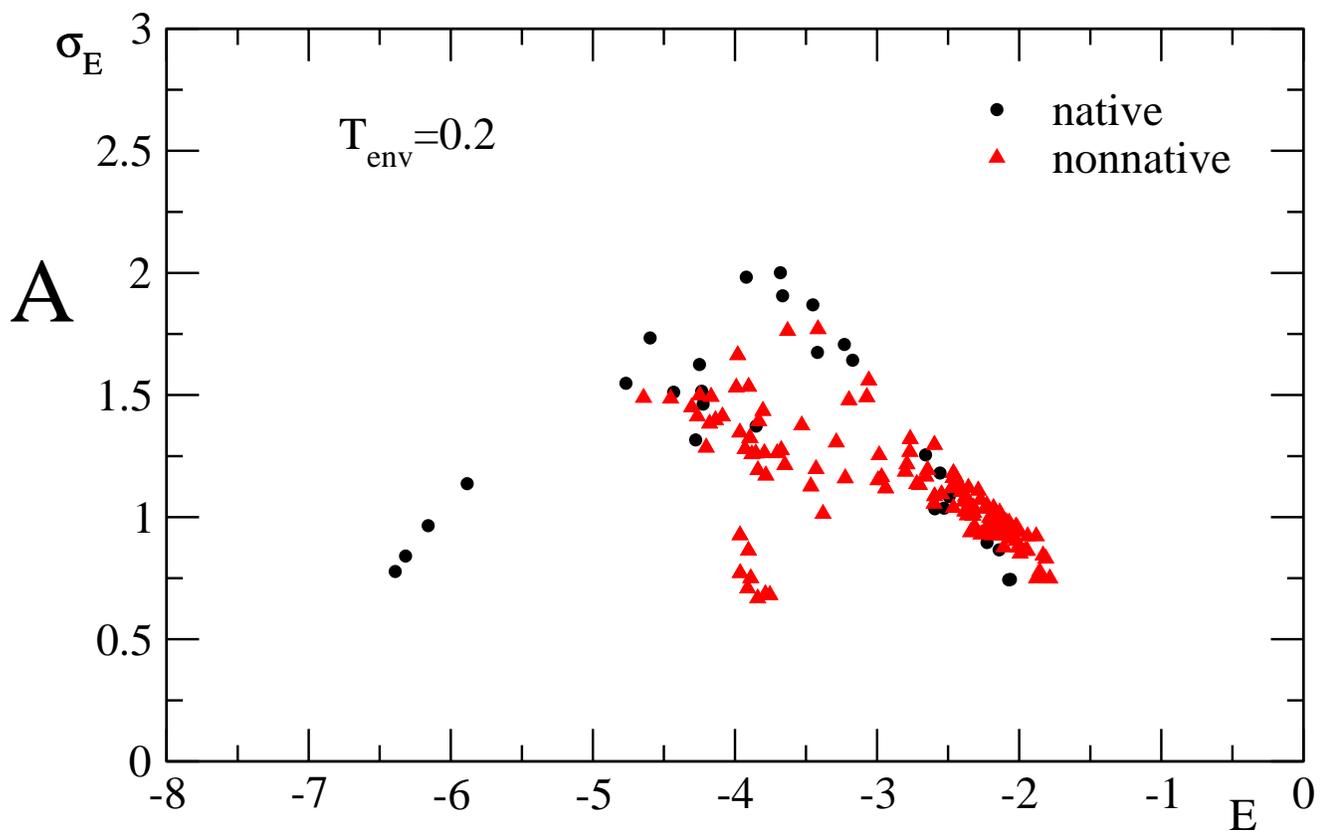

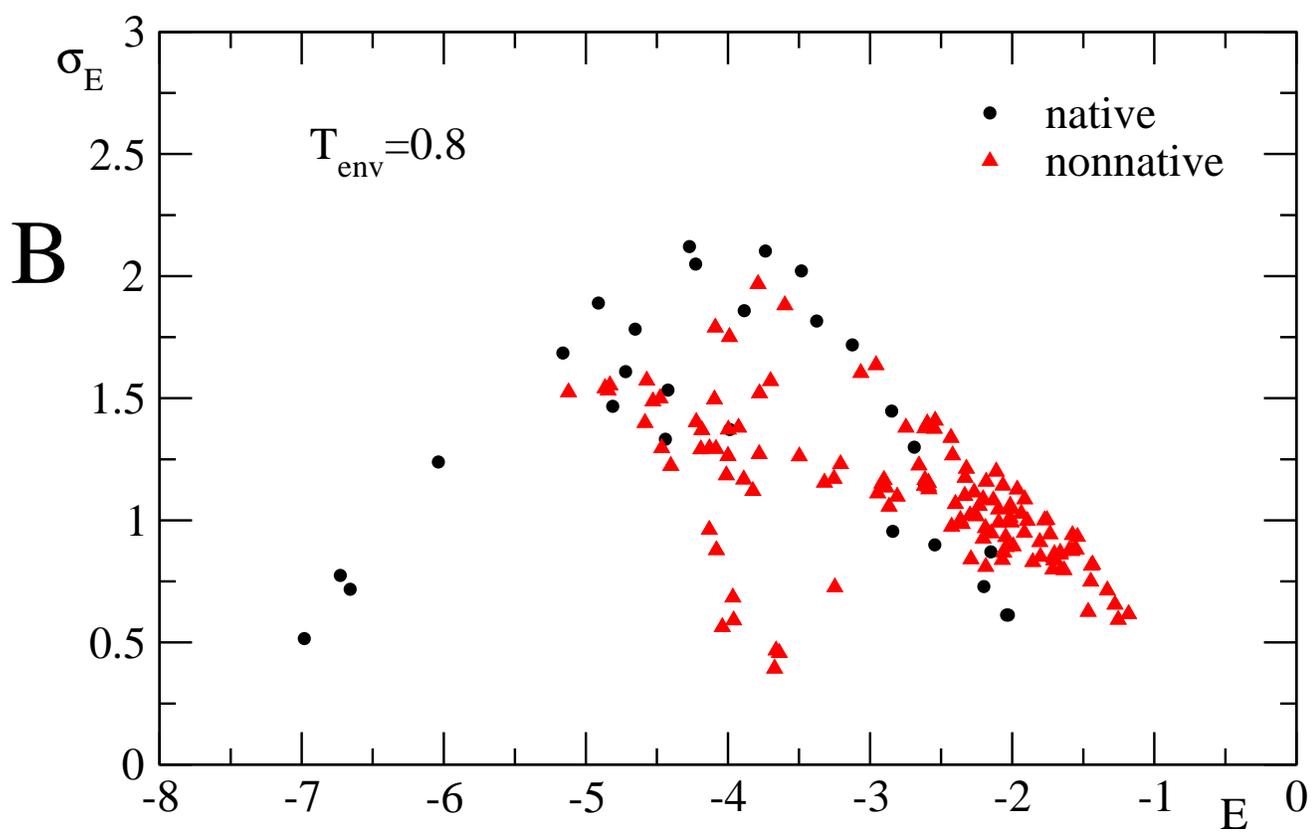

**A**

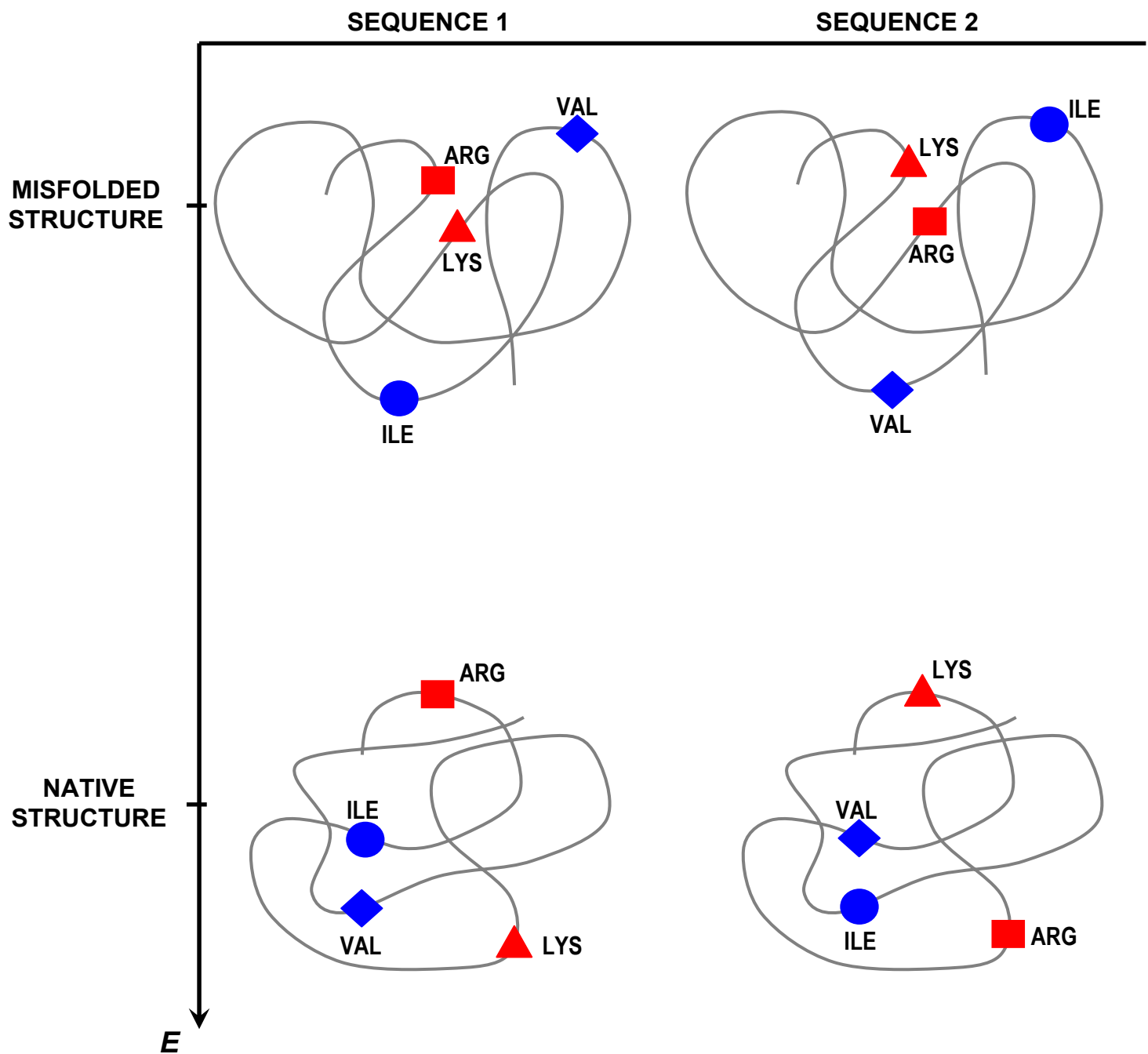

**B**

```
SEQUENCE1:  xxxxxVxxxxxxxKxxxxxxxxxxIxxxxxxxxxRxxxx
SEQUENCE2:  xxxxxIxxxxxxxRxxxxxxxxxxVxxxxxxxxxKxxxx
```

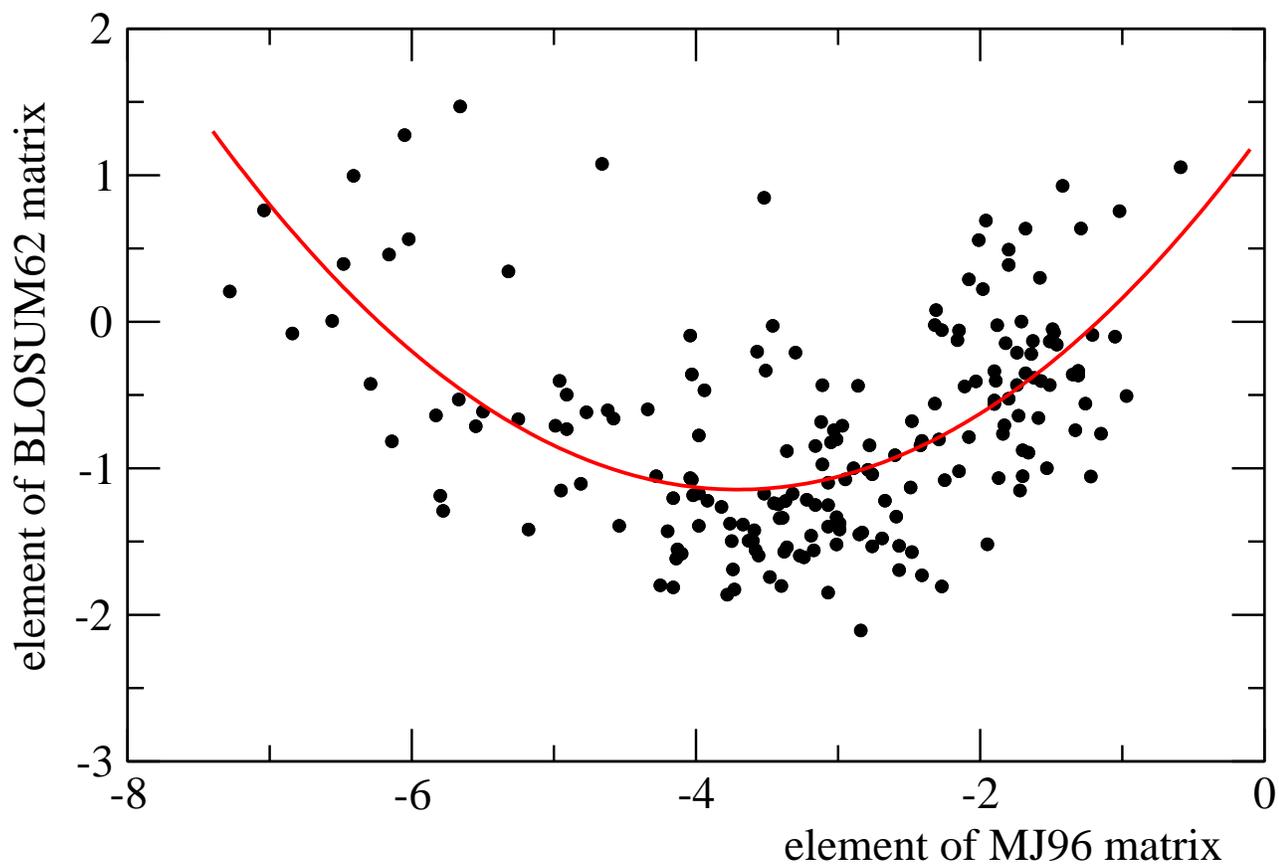

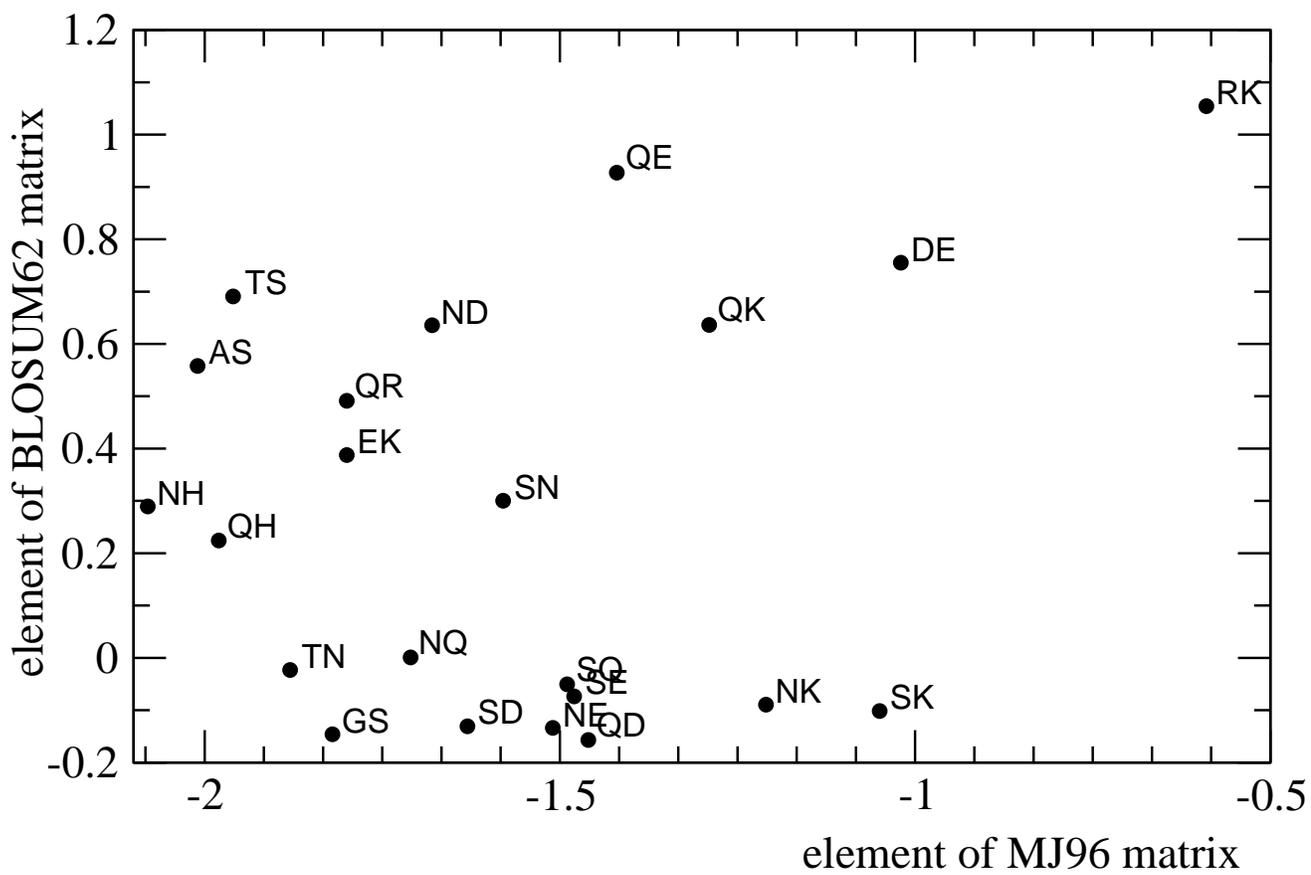

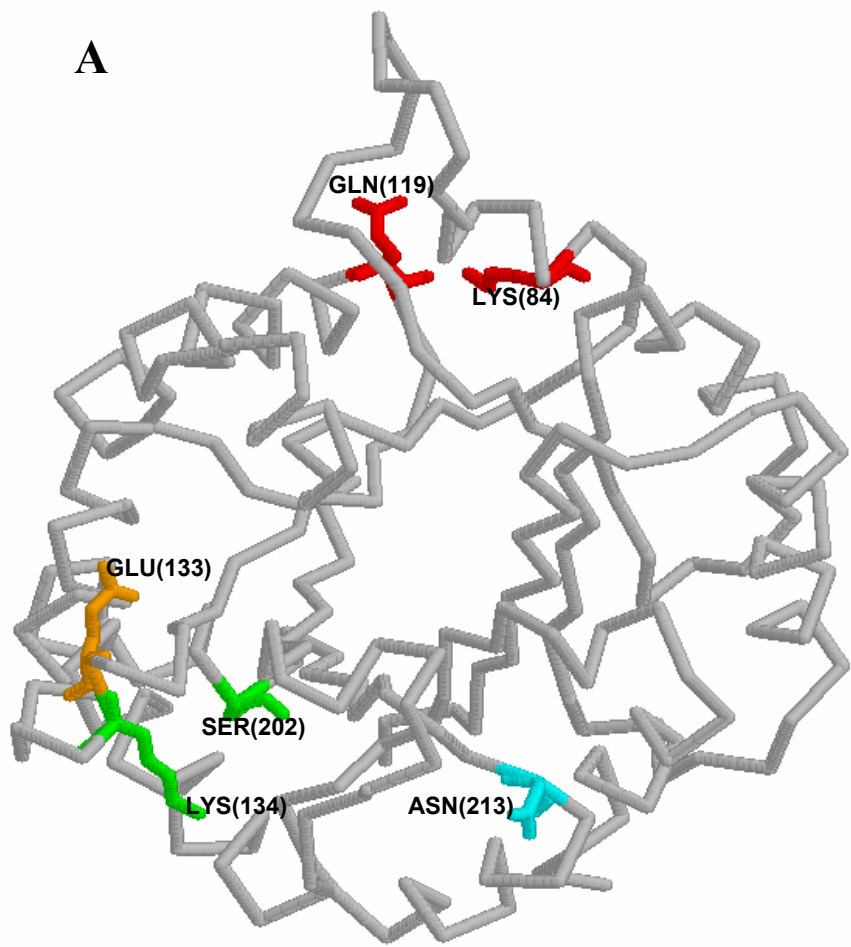
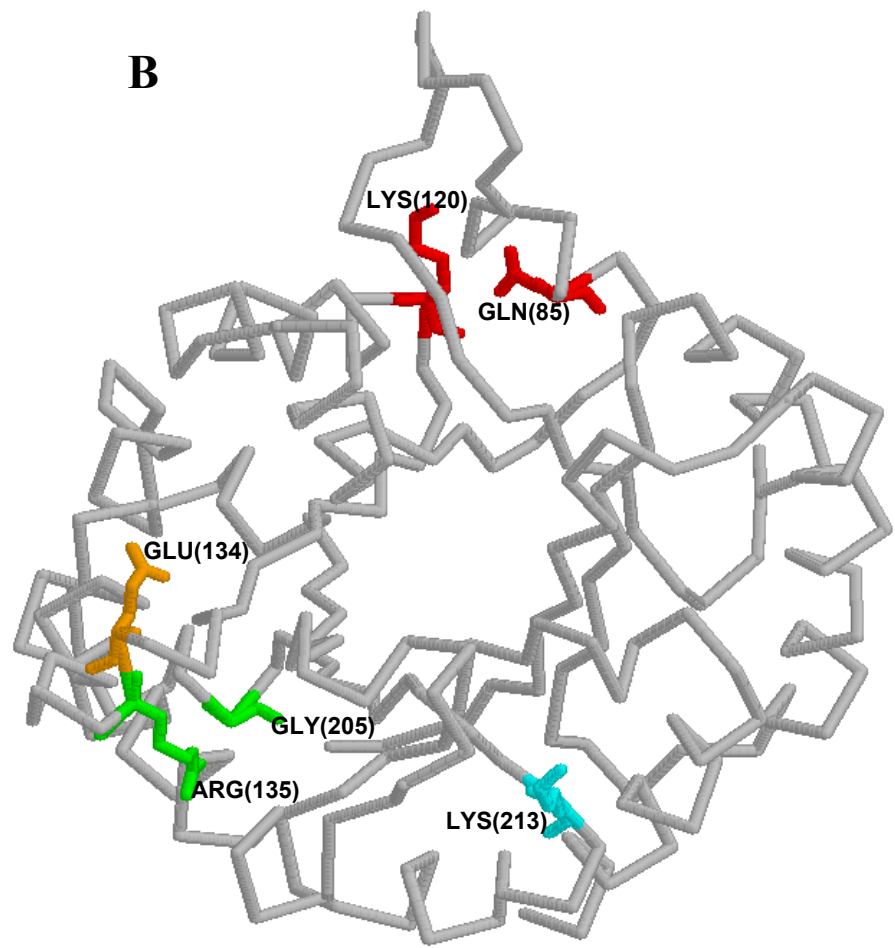